\documentclass[prl,preprint,groupedaddress,showpacs]{revtex4-1}

\usepackage{amsmath}
\usepackage{amssymb}
\usepackage{amsfonts}
\usepackage{mathrsfs}
\usepackage{epsfig}
\usepackage{bm}

\begin{document}

\title{Plasma based helical undulator for controlled emission of circularly and elliptically polarised betatron radiation}
\author{J. Vieira, J. Martins, U. Sinha}
\affiliation{GoLP/Instituto de Plasmas e Fus\~{a}o Nuclear,  Instituto Superior T\'{e}cnico, Universidade de Lisboa, Lisbon, Portugal}

\today

\begin{abstract}
We explore a plasma based analogue of a helical undulator capable of providing circularly and elliptically polarised betatron radiation. We focus on ionisation injection configurations and in the conditions where the laser pulse driver can force collective betatron oscillations over the whole trapped electron bunch. With an analytical model and by employing three dimensional simulations and radiation calculations, we find that circularly or elliptically polarised laser drivers can force helical betatron oscillations, which produce circularly/elliptically polarised betatron x-rays. We assess the level of polarisation numerically and analytically, and find that the number of circularly polarised photons can be controlled by tuning the laser pulse driver polarisation. We show the production of betatron radiation that is circularly polarised up to $\lesssim 40\%$ close to regions of maximum photon flux. The total flux of circularly polarised betatron radiation drops for elliptically polarised drivers, and is negligible when using linear polarisation. Our results can be tested today in current experimental facilities.
\end{abstract}

\maketitle

Plasmas sustain extremely intense fields, orders of magnitude larger than any other material. This unique feature has lead to the development of compact, plasma based light sources. Although there are various radiation emission mechanisms in plasmas addressing specific spectral regions, the production of x- and gamma rays is usually achieved in plasma accelerators~\cite{bib:rousse_prl_2004}. A plasma accelerator uses an intense laser~\cite{bib:tajima_prl_1979} or particle beam driver~\cite{bib:chen_prl_1985} to excite relativistic plasma waves capable of accelerating electron bunches. While accelerating, bunch electrons also perform transverse oscillations (betatron oscillations), which are driven by the focusing fields provided by the background plasma ions. Like in the synchrotron, betatron radiation emission occurs when the acceleration component perpendicular to the velocity is maximum, at the crests of the betatron oscillations~\cite{bib:kostyukov_pop_2003}.

State-of-the-art experiments for the production of betatron radiation operate in the strongly non-linear blowout regime~\cite{bib:pukhov_apb_2002,bib:lu_prl_2006,bib:lu_prstab_2007}. Radiation emission can be significantly enhanced in the blowout when the laser interacts with accelerated electrons, resonantly driving the betatron oscillations~\cite{bib:cippicia_nphys_2011,bib:shaw_ppcf_2014}. State-of-the-art experiments in these regimes typically deliver ultra-fast ($\sim$1-10 fs), spatially collimated ($\lesssim$ 10 mrads), and energetic (10 KeV - 10 MeV) photon bunches~\cite{bib:cippicia_nphys_2011}. These x-rays and gamma-rays, whose properties are fully determined by the details of betatron trajectories~\cite{bib:albert_prl_2013}, can be used to image microscopic structures~\cite{bib:kneip_nphys_2010}, and to generate high resolution tomographic images of medical samples with unprecedented resolutions~\cite{bib:cole_inprep_2015}. In addition to these advances, betatron light sources have the potential to impact an even wider range of applications where imaging plays a vital role, from medicine to nuclear physics~\cite{bib:albert_ppcf_2014}.

In addition to its energy and spatial distribution, it is also important to understand the polarisation properties of betatron radiation. Producing circularly polarised betatron radiation can be particularly interesting. This could open the way to probe the spatial structure of complex molecules (e.g. proteins) and the magnetic properties of materials~\cite{bib:yamamoto_prl_1989} with plasma based betatron radiation. Recent experiments already identified a path to control the production of linearly polarised x-rays in the laser wakefield accelerator~\cite{bib:schnell_ncomms_2013,bib:schnell_jpp_2015} by using a laser with a pulse front tilt~\cite{bib:popp_prl_2010} or in ionisation injection scenarios~\cite{bib:doepp_arxiv_2015}. Similarly to undulator radiation, these x-rays may also contain a circular polarisation component at larger observation angles. However, the number of circularly polarised photons will be negligible in those regions, because most of the energy is concentrated at small angles centred on the propagation direction. As a result, circularly polarised UV, x-ray and gamma-ray radiation can then only be currently produced in helical undulators~\cite{bib:yamamoto_prl_1989}, which use a particular arrangement of magnetic fields to produce helical electron bunch trajectories.

In this Letter we explore a plasma based analogue of the helical undulator, capable of producing and controlling the production of circularly polarised x-rays in a plasma accelerator. Employing theory and three-dimensional OSIRIS~\cite{bib:osiris} simulations, we show that the interaction of a circularly or elliptically polarised laser driver with an electron bunch in ionisation injection scenarios~\cite{bib:pak_prl_2010,bib:mcguffey_prl_2010} leads to helical betatron oscillations of individual bunch particles. By using the appropriate Stokes parameters we then demonstrate, with theory and numerical simulations using the massively parallel radiation code jRad~\cite{bib:jrad}, that these collective helical trajectories can lead to the emission of x-rays with degrees of circular polarisation up to $40\%$. We show that the flux of circularly polarised betatron photons can also be precisely controlled by tuning the laser polarisation, which also determines the spatial distribution of circularly polarised x-ray radiation. Our findings can be tested today with currently available experimental conditions.

%\item{The emission of betatron radiation occurs at the crests of the betatron oscillations, when the focusing fields are maximum, therefore sharing many of the features of synchrotron radiation~\cite{bib:kostyukov_pop_2003}. The majority of betatron radiation experiments operates in the strongly non-linear blowout regime~\cite{bib:pukhov_apb_2002, bib:lu_prl_2006,bib:lu_prstab_2007}, where the betatron radiation yield is maximum. The largest photon energies ever measured were obtained in the blowout regime when betatron oscillations were also driven by the laser pulse driver fields. Experiments in this configuration lead to the emission of gamma rays with energies up to 7 MeVs~\cite{bib:cippicia_nphys_2011}.}

We start by considering the radiation field $\mathbf{E}(t)$ emitted by a moving electron with mass $m_e$ and charge $-e$, given by:
\begin{equation}
\label{eq:erad}
\mathbf{E}(t^{'}) = \left[\frac{e}{c}\frac{\mathbf{n}\times\left\{\left(\mathbf{n}-\boldsymbol{\beta}\right)\times\dot{\boldsymbol{\beta}}\right\}}{(1-\boldsymbol{\beta}\cdot\mathbf{n})^3 R}\right]_{\mathrm{ret}} = \left[\frac{e}{c}\frac{\left(\boldsymbol{\beta} \cdot \mathbf{n}-1\right) \boldsymbol{\dot{\beta}}+\mathbf{n}\cdot\boldsymbol{\dot{\beta}}\left(\mathbf{n}-\boldsymbol{\mathbf{\beta}}\right)}{(1-\boldsymbol{\beta}\cdot\mathbf{n})^3 R}\right]_{\mathrm{ret}}.
\end{equation}
where $c$ is the speed of light, $\mathbf{n}$ is the unit vector that goes from the charge to the observation point, $R(t)=x-\mathbf{n}\cdot \mathbf{r}(t)/c$ is the distance from the charge to the observation point in the detector, $x$ is the distance of the origin to the observation point, $\mathbf{r}(t)$ is the moving charge trajectory, and where $\boldsymbol{\beta}=\mathrm{d}\mathbf{r}/\mathrm{d}t$ and $\dot{\boldsymbol{\beta}}=\mathrm{d}\boldsymbol{\beta}/\mathrm{d}t$ are the velocity and acceleration normalised to c. In addition, the subscript $\mathrm{ret}$ means that quantities are evaluated at the retarded time $t^{'}=t+R(t)/c$. Vector quantities are in bold. 

The right-hand side of Eq.~(\ref{eq:erad}) establishes the relation between the direction of the radiated electric field and the direction of the electron velocity and electron acceleration. In order to clearly evidence this relation we consider the radiation emitted by an ultra-relativistic electron in the far-field. Close to the axis, such that $\mathbf{n}\simeq (0,0,1)\simeq \mathbf{e}_z$ ($\mathbf{e}_z$ is the unit vector pointing in the longitudinal z direction), the transverse electric field components $\mathbf{E}_{\perp}$ given by Eq.~(\ref{eq:erad}) can be re-written as:
\begin{equation}
\label{eq:erad_farfield}
\mathbf{E}_{\perp}(t^{'}) = \left[\frac{e}{c R}\frac{\boldsymbol{\beta}_{\perp} \dot{\beta}_z-\boldsymbol{\dot{\beta}_{\perp}} \left(1-\beta_z\right)}{\left(1-\beta_z\right)^3}\right]_{\mathrm{ret}} 
%= \left[\frac{e}{c R}\frac{1}{{1-\beta_z}}\frac{\mathrm{d}}{\mathrm{d} t} \left(\frac{\boldsymbol{\beta}_{\perp}}{1-\beta_z}\right)\right]_{\mathrm{ret}},
\end{equation}
where $\beta_z$ is the longitudinal electron velocity and where $\boldsymbol{\beta}_{\perp}$ is the transverse electron velocity. Equation~(\ref{eq:erad_farfield}) shows that the polarisation properties can be fully controlled through $\boldsymbol{\beta_{\perp}}$ and through its time derivative $\dot{\boldsymbol{\beta}}_{\perp}$. Thus, radiation will be linearly polarised when the betatron trajectories are defined in a single plane, and circularly or elliptically polarised when electron trajectories are helical. The level of circular polarisation $P_c$ is given by the Stokes parameters, in which $P_c=V/I$, where $V=-2 \langle \mathrm{Im}\left(E_x E_y^{\ast}\right)\rangle$ and where $I=|\mathbf{E}_\perp|^2$. The brackets $\langle \cdot \rangle$ represent a time average needed to describe the polarisation features of light with a broad spectra. We can determine a simple scaling for $P_c$ by assuming simplified betatron trajectories where $\dot{\beta}_z \ll \dot{\beta}_{x}$ or $\dot{\beta}_z \ll \dot{\beta}_{y}$, and by assuming helical betatron trajectories where $\beta_x \sim \exp\left(i \omega_{\beta} t\right)$ and $\beta_y\sim \exp\left(i \omega_{\beta} t + i \varphi\right)$, and where $\varphi$ is a phase. For a single electron executing helical betatron trajectories, $P_c \sim \sin\left(\varphi\right)$. A similar scaling can also be obtained for the angular momentum $L_z\sim \mathbf{r}_{\perp} \times \boldsymbol{\beta}_{\perp}=|\mathbf{r}_{\perp}| |\boldsymbol{\beta}_{\perp}| \sin{\varphi}$. These expressions indicate that the degree of circular polarisation can be controlled by the ellipticity of the trajectory, or, equivalently, by its angular momentum~\cite{bib:thaury_prl_2013}.

The estimate for the level of circular polarisation can be extended for a particle beam with $N$ electrons considering that the transverse velocity of each beam particle is $\beta_{x,n} \sim \cos\left(\omega_{\beta,n} t\right)$ and $\beta_y\sim \cos\left(\omega_{\beta,n} t + \varphi_n\right)$. Neglecting effects associated with the longitudinal acceleration (terms proportional to $\dot{\beta}_z$ in Eq.~(\ref{eq:erad_farfield})), the resulting electric field is $\mathbf{E}_\perp \propto \Sigma_n \boldsymbol{\beta}_n$. Employing the random phase approximation~\cite{bib:silva_pre_1999}, and assuming that the retarded time is similar for all electrons except for a constant factor associated with its initial position, then gives $P_c \sim \langle \mathrm{Im}\left( \Sigma_{n,m} \beta_{x,n} \beta_{y,m}^{\ast}\right)\rangle \sim  \langle Im\left\{ \sum_{n,m} \exp\left[i\left(\omega_m-\omega_n\right) t-i \varphi_m\right]\right\}\rangle \sim (1/N) \sum_n\sin{\varphi_n}$. In practice, the latter assumption implies that electrons are characterised by a same relativistic factor $\gamma$. The total amount of circular polarisation then corresponds to the average circular polarisation value of each electron. As a result, the production of circularly polarised betatron x-rays can be achieved when each single beam electron performs a helical betatron trajectory.

There are various possibilities to generate helical betatron oscillations in a plasma accelerator. Examples include injecting an electron bunch off axis~\cite{bib:glinec_epl_2008,bib:vieira_prl_2011} with a transverse velocity component, using external longitudinal magnetic fields; employing asymmetric drivers that create transversely evolving plasma bubbles~\cite{bib:thaury_prl_2013}, and forcing betatron oscillations in the presence of laser fields~\cite{bib:nemeth_prl_2008}. Here we explore a mechanism to produce and control the production of circularly polarised betatron x-rays when electron betatron trajectories are forced by a laser pulse driver with various polarisations (linear, elliptical or circular) in an ionisation injection scenario~\cite{bib:pak_prl_2010,bib:mcguffey_prl_2010}, recently also considered as a suitable candidate to produce and control the emission of linearly polarised x-rays~\cite{bib:doepp_arxiv_2015}.

We consider a laser pulse driver with normalised vector potential given by $e \mathbf{A}_{\mathrm{\perp}}/(m_e c) = a_0 f(z-v_g t) \left[\cos\left(k_0 z - \omega_0 t\right),\cos\left(k_0 z - \omega_0 t + \varphi\right)\right]$, where  $f(z-v_g t)$ is the laser longitudinal profile, $v_g$ is the linear laser group velocity, $k_0$ and $\omega_0$ the laser central wavenumber and frequency respectively, and $\varphi$ is a constant phase that determines the laser polarisation (linear polarisation for $\varphi=0$, circular polarisation for $\varphi=\pm\pi/2$, and elliptical polarisation for other values of $\varphi$). The laser wavenumber and frequency are related through the linear plasma dispersion relation in an underdense plasma, given by $\omega_0 \simeq k_0 c \left[1-\omega_p^2/(2 \omega_0^2)\right]$, where $\omega_p$ is the plasma frequency. When $\gamma_{z} \gg \gamma_g$, where $\gamma_g \simeq \omega_0/\omega_p $ is the relativistic factor associated with the linear laser group velocity, $\gamma_{z}=1/(1-v_z^2/c^2)$ is the Lorentz factor of the electrons, and for a smooth envelope profile such that $k_0 \mathrm{d}f(\xi) /\mathrm{d} \xi \ll 1$, the equation for the transverse electron motion in the blowout regime in the co-moving frame $\xi=z-v_g t$ becomes:
\begin{equation}
\label{eq:eom}
 \frac{\mathrm{d}^2 \mathbf{x}_{\perp}}{\mathrm{d} \xi^2} + \frac{\mathrm{d}\gamma}{\mathrm{d}\xi}\frac{\mathrm{d}\mathbf{x}_{\perp}}{\mathrm{d}\xi} + 4 \gamma_{g}^2 \omega_{\beta}^2 \mathbf{x}_{\perp} = 2 a_0 k_0 \gamma_g^2 f(\xi) \sin\left[k_0 \xi \left(1+\frac{2 \gamma_g^2}{\gamma_{\|}^2}\right) + \boldsymbol{\varphi} \right].
\end{equation}
Equation~(\ref{eq:eom}) recovers the work of Ref.~\cite{bib:nemeth_prl_2008} with $\boldsymbol{\varphi}=(0,\varphi)=(0,0)$. Equation~(\ref{eq:eom}) describes an harmonic oscillator, with natural frequency corresponding to the doppler shifted betatron frequency ($2 \gamma_g \omega_{\beta}$), with damping ($\propto \mathrm{d}\gamma/\mathrm{d} \xi$) and with a driving (right hand side of Eq.~(\ref{eq:eom})) term. The damping term suppresses pure betatron oscillations, being the dominant contribution at early times~\cite{bib:nemeth_prl_2008}. Then, the motion reaches a steady state and betatron trajectories become solely driven by the laser pulse. The steady-state is reached after $\xi_{\mathrm{s}}\gtrsim 2 (\mathrm{d}\gamma / \mathrm{d} \xi)^{-1}$. Since $\mathrm{d}\gamma/\mathrm{d}\xi \simeq E_{\mathrm{accel}}$, where $E_{\mathrm{accel}}\simeq \sqrt{a_0}$ is the average accelerating gradient, and since for an electron traveling at nearly the speed of light with $v_z\simeq c$, the steady state $t_{\mathrm{s}}$ is reached after $t_{\mathrm{s}} = 2 \xi_{\mathrm{s}} \gamma_g^2\simeq 4 \gamma_g^2/ \sqrt{a_0}$. For $\xi>\xi_{\mathrm{s}}$, electron trajectories become given by:
\begin{equation}
\label{eq:helical}
\mathbf{x}_{\perp} \propto \sin\left[k_0 \xi\left(1+\frac{2 \gamma_g^2}{\gamma_{\|}^2} \right) + \boldsymbol{\varphi} + \phi\right],
\end{equation}
where $\phi$ is a phase, identical for all particles at each $\xi$. Hence, Eq.~(\ref{eq:helical}) shows that the bunch and individual bunch particles execute coherent oscillations both in $\xi$ and in $t$ [because $\xi$ maps time $t$ through $\xi\simeq (v_{\|}-v_{g}) t$] for $\xi\gtrsim \xi_{s}$. The trajectories are planar when the driver is linearly polarised ($\varphi = 0$). Hence, $P_c=0$ when $\varphi=0$. When the laser is circularly polarised ($\varphi  = \pm \pi/2$) the trajectories are helical, thereby maximising $|P_c|=1$. For other values of $\varphi$, radiation is elliptically polarised with $0<|P_c|<1$.

We have confirmed these predictions by post-processing the particle trajectories of three-dimensional (3D) Osiris~\cite{bib:osiris} simulations employing the radiation code jRad~\cite{bib:jrad}, which predicts the spatially resolved radiation spectrum and polarisation. We explored the production and control of circularly polarised betatron x-rays in ionisation injection scenarios, considering parameters that are available in many laboratories. We then used parameters close to those of Ref.~\cite{bib:pak_prl_2010}. The laser pulse has normalised vector potential $a_0 = 1.7$, transverse spot-size $w_0 = 6.6~\mathrm{\mu m}$ (1/e) and full width half maximum duration (fields) $\tau_{\mathrm{FWHM}}=57~\mathrm{fs}$. It propagates into a gas with a mixture of Helium, with density $n_{\mathrm{He}}=4\times10^{18}~\mathrm{cm}^{-3}$, and Nitrogen, with density $n_{\mathrm{N}} = 2\times10^{16}~\mathrm{cm}^{-3}$. The gas density rises for $50 c/\omega_p$ ($\simeq 100~\mu m$), is flat during $500 c/\omega_p$ ($\simeq 1~\mathrm{mm}$) and falls back to zero for another $50 c/\omega_p$ ($\simeq 100~\mu m$).

\begin{figure}
\centering\includegraphics[width=0.6 \columnwidth]{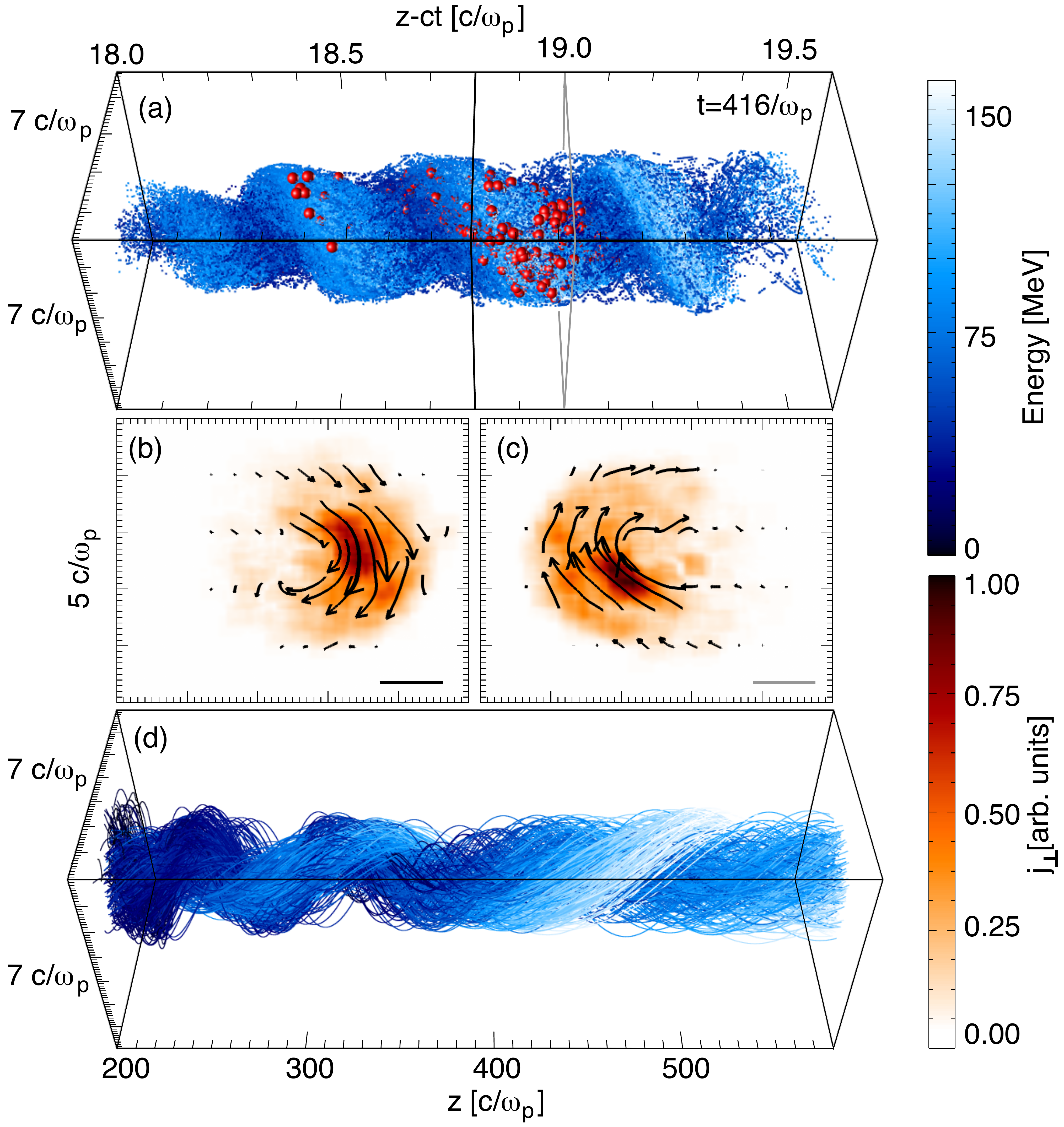}
\caption{Osiris simulation results illustrating forced betatron oscillations of an ionisation injected electron bunch (outer shell of Nitrogen) interacting with the laser driver. (a) shows the 3D electron trapped electron bunch distribution at $t=507/\omega_p$. Each electron is coloured in blue according to its energy. (b) and (c) show the transverse particle momentum vector plot at the longitudinal positions marked by the grey and black squared lines in (a). (d) shows the trajectories of a selected group of ionisation injected particles indicated by the red spheres in (a). }
\label{fig:edynamics}
\end{figure}

We begin by studying simulations where $\varphi=-\pi/2$. The laser fully ionises the Helium gas and the inner (1-5) Nitrogen electron shells at the entrance of the plasma. At end of the up-ramp, the laser drives non-linear plasma waves in the blowout regime. Trapping and acceleration of electrons from the outer (6-7) Nitrogen shells also occur at the end of the initial plasma ramp. These electrons, which gain up to 150 MeVs at the end of the plasma (and before dephasing), begin interacting with the laser pulse driver shortly after being trapped. The laser ionisation rates for the outer Nitrogen shells forms a helical pattern closely following the laser electric field direction. As the laser propagates, it keeps ionising the outer Nitrogen shells, therefore smoothing the helical structures. Simultaneously, the laser also drives the betatron oscillations, which according to Eq.~(\ref{eq:helical}), lead to a whole bunch modulation with a periodicity close to the laser wavelength. These modulations, which are clear from Fig.~\ref{fig:edynamics}a, correspond to a helical structure mapping the laser pulse field polarisation directions.

Figures~\ref{fig:edynamics}(b)-(c), which illustrate transverse velocity vector plots at two different positions within the bunch, demonstrate that bunch electrons acquired an azimuthal velocity component in the anti-clockwise direction, performing global helical trajectories. Additional simulations with $\varphi = \pi/2$ show that electrons rotate clockwise instead. The collective motion of each bunch slice, containing electrons from close longitudinal positions, also follows a helical path as shown in Fig.~\ref{fig:edynamics}(d). 

%\item{The global helical motion of the whole electron bunch is critical to enhance the degree of circular polarisation of the betatron x-rays. A bunch with uniformly distributed density (e.g. cylindrically symmetric bunch), where each electron performs an helical trajectory. Near the center the field of each electron is exactly canceled by the field of another electron propagating at $\pi$ out of phase with respect to the first. In reality, this cancelation is only partial, but the direction of the field may change randomly.}
%\begin{figure}
%\centering\includegraphics[width=\columnwidth]{figures/figure1.pdf}
%\caption{Osiris simulation results showing transverse slices of the electron plasma density (gray), laser pulse electric fields (rainbow) and ionisation injected electrons (blue) at $t=156/\omega_p$ (a) and at $t=429/\omega_p$ (b). (c) shows details of the self-injected bunch superimposed with the laser fields in the region delimited by the dotted lines in (b).}
%\label{fig:ioninj}
%\end{figure}

These features suggest that electrons are producing circularly polarised x-rays. In order to evaluate $P_c$ we then performed jRad radiation calculations considering various sets of trajectories of 512 particles from the outer Nitrogen shell electrons. Radiation was retrieved in a virtual detector located at $z=10^4 c/\omega_p$, far from the exit of the plasma. The detector provides spatially resolved energy and polarisation spectra. For all configurations explored, the detector is $1000 (c/\omega_p) \times 1000 (c/\omega_p)$ divided into $50\times50$ cells. The detector captures frequencies up to $\omega_{\mathrm{max}}=6\times10^4 \omega_p$ being divided into $2^{15}$ (32168) cells in the frequency axis for all cases. Radiation calculations refer to the range starting with particle injection at $t_i=130/\omega_p$ up to the end of the plasma at $t_f=650/\omega_p$.

\begin{figure}
\centering\includegraphics[width=0.8 \columnwidth]{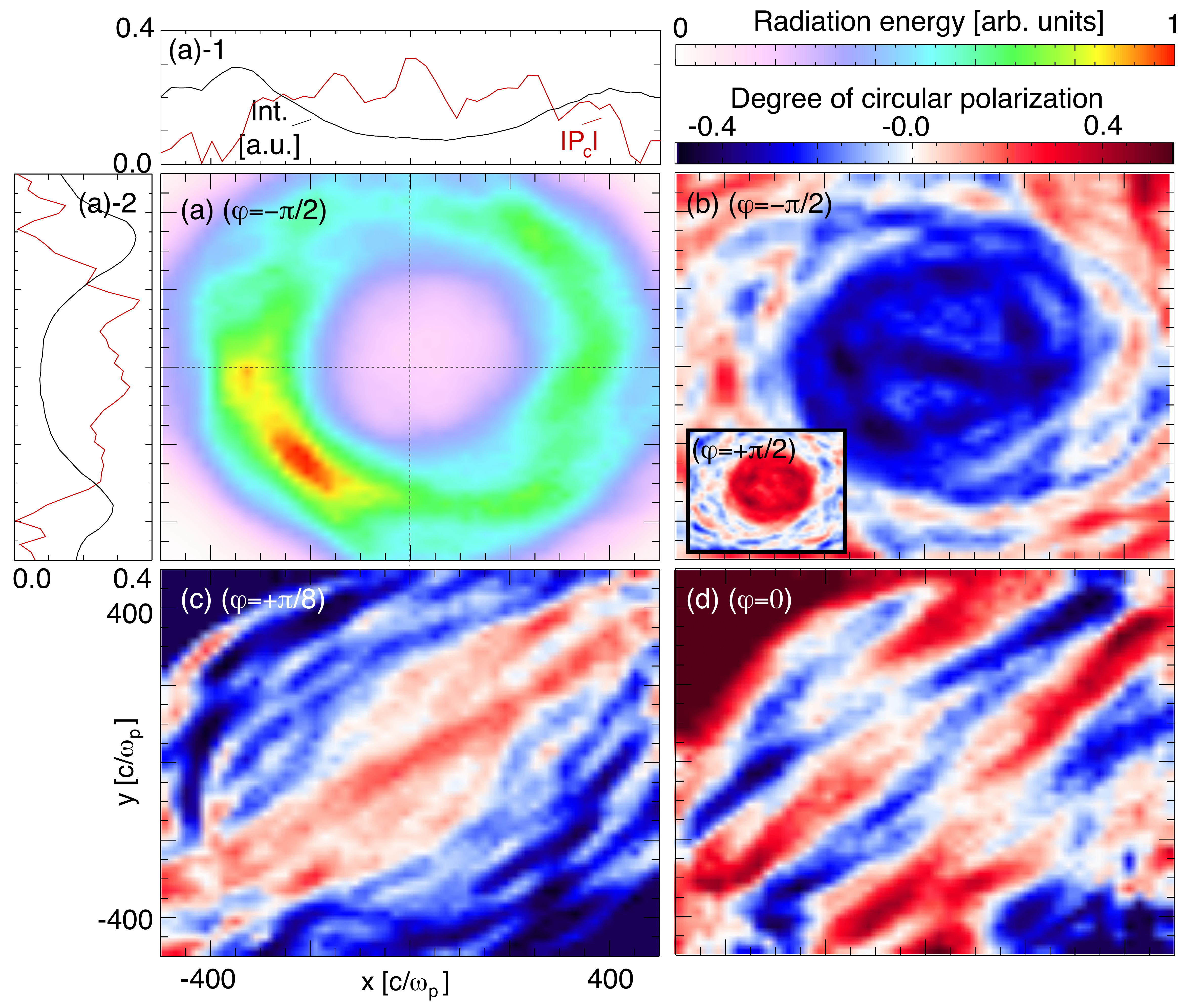}
\caption{jRad simulation results illustrating key properties of the radiation emission in ionisation injection scenarios. (a) shows the x-ray energy profile integrated in $\omega$ for the group of electron trajectories of Fig.~\ref{fig:edynamics}d. (b) shows the corresponding average level of circular polarisation integrated in $\omega$. (c)-(d) show $P_c$ for a elliptically ($\varphi=\pi/4$) and linearly ($\varphi=0$) polarised laser drivers.}
\label{fig:pc}
\end{figure}

Figure~\ref{fig:pc} shows the radiated energy spectrum ($I_{\mathrm{rad}}$) integrated for frequencies $\omega > 10^3~\omega_p$ (well above ultraviolet) for the same trajectories in Fig.~\ref{fig:edynamics}(d) with $\varphi=-\pi/2$. Since bunch electrons perform spiralling trajectories, the electron bunch density profile has a minimum on axis. As a result, most of the radiation is emitted off axis at an angle $\theta\sim K/\gamma$. As the angular width around the emission angle $\Delta \theta \sim 1/\gamma$ is smaller than $\theta$, the transverse radiation profile acquires the doughnut shape seen in Fig.~\ref{fig:pc}a. The normalised, average flux of circularly polarised photons $\langle P_c\rangle = \int P_c I_{\mathrm{rad}} \mathrm{d}\omega/ \int I_{\mathrm{rad}} \mathrm{d}\omega$ is shown in Fig.~\ref{fig:pc}b, indicating the production of x-rays with peak $P_c \sim 35\%$. This is lower than theoretical predictions, for which $P_c=-1$ when $\varphi = - \pi/2$. The main reason for the discrepancy is that the transverse electron trajectories in simulations are not purely circular, being elliptical instead. According to the theoretical model, resonant betatron oscillations start when $t\gtrsim t_{\mathrm{s}} \simeq 600 / \omega_p$. As a result, $P_c$ could be enhanced for longer propagation distances, when the betatron resonance becomes stronger. Additional simulations for longer propagation distances confirm higher degrees of circular polarisation in excess of $40\%$.

We note that our jRad calculations assume that light is fully polarised. In order to confirm that this assumption is not changing our conclusions, we performed time averaged calculations by splitting the particle trajectories in intervals that start at the same propagation distance but ending at different distances. The degree of circular polarisation was identical in all intervals. As a result, the average $P_c$ value is similar to the results we show here. We have further confirmed the validity of our calculations by considering different random sets of particles. In all cases, the level of circular polarisation did not vary much. These results confirm that our predictions for $P_c$ are physically meaningful.

%For elliptically or linearly polarised laser drivers, simulation show the production of a braided electron bunch, as ionisation takes place around the angles where the laser fields are maximum. These beams emit circularly polarised x-rays and the number of photons with circular polarisation can be controlled through $\varphi$.

Key circular polarisation features such as its handedness and spatial distribution can be controlled by varying $\varphi$. For instance, when $\varphi=\pi/2$ (inset of Fig.~\ref{fig:pc}b), $P_c>0$ reversing its handedness, in agreement with theoretical predictions. When $\varphi = \pi/8$, the maximum value of $P_c$ lowers, and the region where $P_c>0$ follows an ellipsoidal shape that coincides with the region where the betatron radiation energy distribution is maximum. We note that the circular polarisation can change its handedness at larger angles. The handedness change, however, occurs in regions where the radiated energy is smaller, having only a minor influence on the flux of circularly polarised photons.  When $\varphi = 0$, the circular polarisation pattern becomes disordered, such that the $P_c$ value averaged over the full detector is negligible (Fig.~\ref{fig:pc}d). Furthermore, $P_c\simeq 0$ for $\varphi = 0$ (Fig. \ref{fig:pc}d), as electrons perform trajectories in a plane (defined by the laser polarisation). We note that when $\varphi=0$, and similarly to a planar undulator, $|P_c|\lesssim 1$ at larger angles. However, the number of photons in those regions is negligible. Fig.~\ref{fig:pc} then demonstrates that the handedness and spatial distribution of $P_c$ can be controlled through $\varphi$.

\begin{figure}
\centering\includegraphics[width=0.8 \columnwidth]{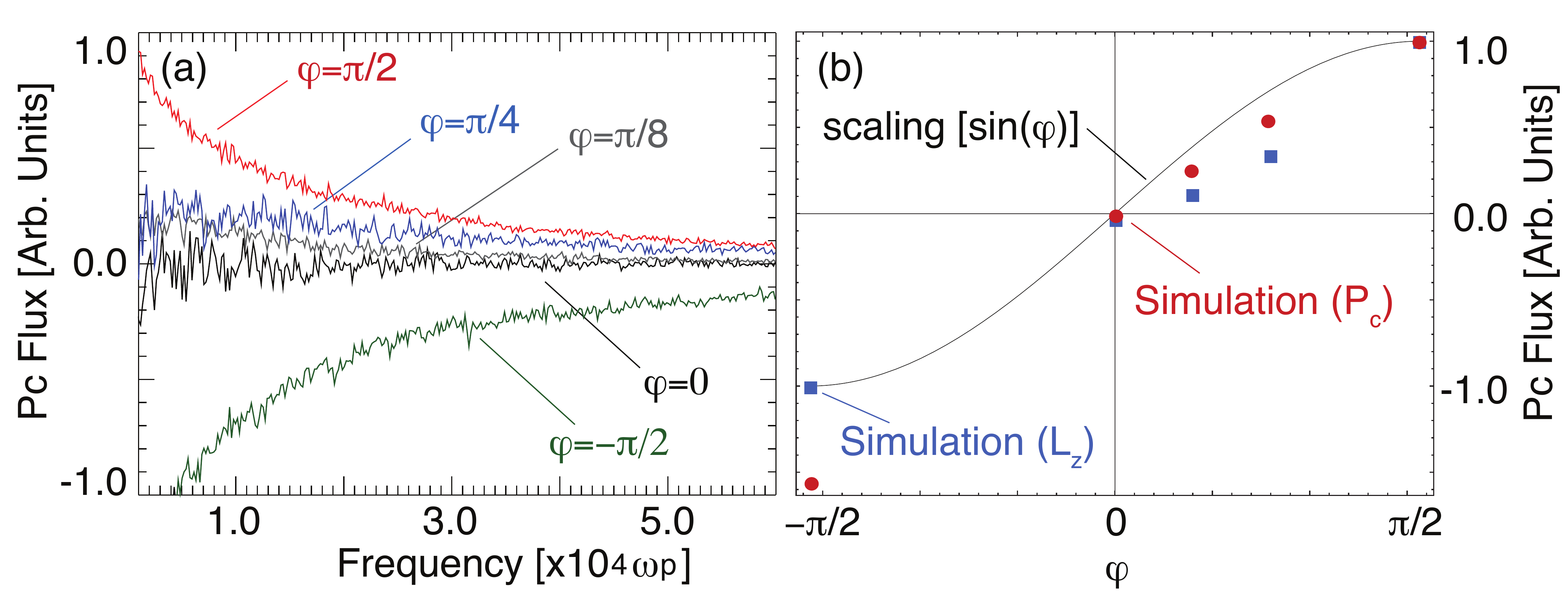}
\caption{jRad simulation results illustrating that the flux of circularly polarised betatron photons can be controlled by the laser pulse driver polarisation. (a) shows the average circular polarisation flux using lasers with various polarisations from $\varphi=0$ to $\varphi=\pi/2$. (b) shows the total circular polarisation flux from simulations (red dots) and the theoretical scaling for $P_c$ (solid line) as a function of the laser driver polarisation.}
\label{fig:polcontrol}
\end{figure}

The flux of circularly polarised photons can also be controlled through $\varphi$. Figure~\ref{fig:polcontrol} shows the total flux of circularly polarised radiation, given by $\mathcal{F}(\omega) = \int P_c E_{\mathrm{rad}} d\mathbf{x}_{\perp} = \int P_c \mathbf{E}^2_{\perp} d\mathbf{x}_{\perp}$ (integration performed over the entire detector) for various $\varphi$. Figure~\ref{fig:polcontrol}a, which illustrates $\mathcal{F}(\omega)$ indicates that the flux of circularly polarised photons is nearly zero using a linearly polarised laser with $\varphi=0$, and maximum for circularly polarised driver with $\varphi=\pm \pi/2$. Figure~\ref{fig:polcontrol}b, which illustrates $\int \mathcal{F}(\omega) \mathrm{d}\omega$, further confirms that the number of circularly polarised photons can be controlled through $\varphi$. Figure~\ref{fig:polcontrol}b also shows that the total flux of circularly polarised x-rays also follows the variations of the total angular momentum of the beam as a function of the laser polarisation.

Although we have explored a particular configuration leading the production of circularly polarised x-rays in plasma based accelerators, we note that many others exist. For instance, ionisation injection scenarios using shorter lasers, where electrons do not interact with the laser while accelerating, could also lead to the production of circularly polarised radiation because electrons are born with a transverse momentum in the direction of the laser electric field thereby causing helical trajectories. The propagation of non-Gaussian lasers with elliptical cross-sections leads to oscillations of the focusing force that induce helical betatron trajectories and circularly polarised x-rays~\cite{bib:thaury_prl_2013}. External longitudinal magnetic fields or external injection at an angle with respect to the propagation axis could also be used to force helical oscillations capable of producing circularly polarised radiation. 

The longitudinal acceleration can change the level of circular polarisation. Our simulations show that the circular polarisation of individual particles can be nearly 100 \% even when their trajectories in the transverse plane for an ellipse. For the trajectories we analysed, this degree of circular polarisation can much larger than our predicted value for a given beam ellipticity when neglecting the term related to the longitudinal acceleration in Eq.~(\ref{eq:erad_farfield}). Because particles emit radiation that arrives to the detector with different phases, the total circular polarisation of the full beam is lower than that of individual particles. However, we have observed very high polarisations close to 100$\%$ when selecting particles within a very narrow energy spread. This suggests that mechanisms capable of strongly reducing the energy spreads in plasma accelerators could significantly enhance the degree of circular polarisation and the number of photons with circular polarisation. The role of longitudinal acceleration will be examined in a future work. 

This scheme can be tested in currently available experimental facilities, and may open new paths for betatron radiation studies and applications. Our results may also provide additional diagnostics of direct laser acceleration in plasma accelerators. 

\acknowledgements

Work supported by the European Research Council (ERC-2010-AdG Grant 267841). We acknowledge PRACE from awarding access to resource SuperMUC (Leibniz research center). The authors also acknowledge fruitful discussions with Prof. Luis Silva and Prof. Ricardo Fonseca. 

%In conclusion, we have explored a mechanism capable of controlling the production of circularly polarised betatron radiation in plasma accelerators. We showed that $P_c$ can be controlled effectively in ionisation injection scenarios according to the laser pulse driver polarisation. These mechanisms may be combined with other setups in order to achieve further control over the production of circularly polarised betatron radiation, for instance, through the use of external magnetic fields or external injection. Extending the propagation towards dephasing in order to achieve resonant betatron oscillations may also increase the level of circular polarisation. 


\begin{thebibliography}{30}

\bibitem{bib:rousse_prl_2004} A. Rousse, K.T. Phuoc, R. Shah, A. Pukhov, E. Lefebvre, V. Malka, S. Kiselev, F. Burgy, J.-P. Rousseau, D. Umstadter, D. Hulin, Phys. Rev. Lett. \textbf{93} 135005 (2004).
%\bibitem{bib:vranic_prl_2014} M. Vranic, J.L. Martins, J. Vieira, R.A. Fonseca, L.O. Silva, Phys. Rev. Lett. \textbf{113}, 134801 (2014).
\bibitem{bib:tajima_prl_1979} T. Tajima, J.M. Dawson, Phys. Rev. Lett. \textbf{43}, 267 (1979).
\bibitem{bib:chen_prl_1985} P. Chen, J. M. Dawson, R. W. Huff and T. Katsouleas, Phys. Rev. Lett. \textbf{54}, 693 (1985).
\bibitem{bib:kostyukov_pop_2003} I. Kostyukov, S. Kiselev, A. Pukhov, Phys. Plasmas \textbf{10} 4818 (2003).
\bibitem{bib:pukhov_apb_2002} A. Pukhov and J. Meyer ter Vehn, Appl. Phys. B \textbf{74}, 355 (2002).
\bibitem{bib:lu_prl_2006} W. Lu, C. Huang, M. Zhou, W. B. Mori, and T. Katsouleas, Phys. Rev. Lett. \textbf{96}, 165002 (2006).
\bibitem{bib:lu_prstab_2007} W. Lu, M. Tzoufras, C. Joshi, F. S. Tsung, W. B. Mori, J. Vieira, R. A. Fonseca, and L. O. Silva, Phys. Rev. ST Accel. Beams \textbf{10}, 061301 (2007).
\bibitem{bib:cippicia_nphys_2011} S. Cipiccia,	M. R. Islam, B. Ersfeld, R. P. Shanks, E. Brunetti, G. Vieux, X. Yang, R. C. Issac, S. M. Wiggins, G. H. Welsh, M.-P. Anania, D. Maneuski, R. Montgomery, G. Smith, M. Hoek,	D. J. Hamilton,	N. R. C. Lemos,	D. Symes, P. P. Rajeev, V. O. Shea, J. M. Dias, D. A. Jaroszynski, Nat. Phys. \textbf{7}, 867 (2011).
\bibitem{bib:shaw_ppcf_2014} J.L. Shaw, F.S. Tsung, N. Najafabi-Vafaei, K.A. Marsh, N. Lemos, W.B. Mori, C. Joshi, Plasma Phys. Control. Fusion \textbf{56}, 084006 (2014). 
\bibitem{bib:albert_prl_2013} F. Albert, B. B. Pollock, J. L. Shaw, K. A. Marsh, J. E. Ralph, Y.-H. Chen, D. Alessi, A. Pak, C. E. Clayton, S. H. Glenzer, and C. Joshi, Phys. Rev. Lett. \textbf{111}, 235004 (2013).
\bibitem{bib:kneip_nphys_2010} S. Kneip, C. McGuffey, J. L. Martins, S. F. Martins,	C. Bellei, V. Chvykov, F. Dollar, R. Fonseca, C. Huntington, G. Kalintchenko, A. Maksimchuk, S. P. D. Mangles, T. Matsuoka, S. R. Nagel, C. A. J. Palmer,	J. Schreiber, K. Ta Phuoc, A. G. R. Thomas,	V. Yanovsky, L. O. Silva, K. Krushelnick, Z. Najmudin, Nat. Physics \textbf{6} 980 (2010).
\bibitem{bib:cole_inprep_2015} J. Cole, N. Lopes, Z. Najmudin \emph{private communication} (2015).
\bibitem{bib:albert_ppcf_2014} F. Albert, A. G. R. Thomas, S. P. D. Mangles, S. Banerjee, S. Corde, A. Flacco, M. Litos, D. Neely, J. Vieira, Z. Najmudin, R. Bingham, C. Joshi, T Katsouleas Plasma Phys. Control. Fusion \textbf{56} 084015 (2014).
\bibitem{bib:yamamoto_prl_1989} S. Yamamoto, H. Kawata, H. Kitamura, M. Ando, N. Sakai, N. Shiotani, Phys. Rev. Lett. \textbf{62} 2672 (1989).
\bibitem{bib:schnell_ncomms_2013} M. Schnell,	A. S{\"a}vert, I. Uschmann,	M. Reuter, M. Nicolai, T. K{\"a}mpfer, B. Landgraf,	O. J{\"a}ckel, O. Jansen,	A. Pukhov,	M. C. Kaluza, C. Spielmann, Nat. Comms \textbf{4} 2421 (2013).
\bibitem{bib:schnell_jpp_2015} M. Schnell,	A. S{\"a}vert, I. Uschmann, O. Jansen, M. C. Kaluza, C. Spielmann, J. Plasma Phys. \textbf{81} 475810401 (2015).	
\bibitem{bib:popp_prl_2010} A. Popp, J. Vieira, J. Osterhoff, Zs. Major, R. Horlein, M. Fuchs, R. Weingartner, T. P. Rowlands-Rees, M. Marti, R. A. Fonseca, S. F. Martins, L. O. Silva, S. M. Hooker, F. Krausz, F. Gruner, and S. Karsch, Phys. Rev. Lett. \textbf{105} 215001 (2010).
\bibitem{bib:doepp_arxiv_2015} A. Doepp, B. Mahieu, A. Doche, C. Thaury, E. Guillaume, A. Lifschitz, G. Grittani, O. Lund, M. Hansson, J. Gautier, M. Kozlova, J. P. Goddet, P. Rousseau, A. Tafzi, V. Malka, A. Rousse, S. Corde, K. Ta Phuoc, arXiv:1509.08629 [physics.plasm-ph] (2015).
\bibitem{bib:osiris} R.A.Fonseca, L.O.Silva, F.S.Tsung, V.K.Decyk, W.Lu, C.Ren, W.B.Mori, S.Deng, S.Lee, T.Katsouleas, and J.C.Adam, Lect. Notes Comp. Sci. vol. 2331/2002, (Springer Berlin / Heidelberg,(2002); R.A. Fonseca, J. Vieira, F. Fi\'uza, A. Davidson, F.S. Tsung, W.B. Mori, L.O. Silva, Plasma Phys. Control. Fusion, 55 124011 (2013).
\bibitem{bib:pak_prl_2010} A. Pak~\emph{et al.}, Phys. Rev. Lett. \textbf{104}, 025003 (2010).
\bibitem{bib:mcguffey_prl_2010} C. McGuffey, A.G.R. Thomas, W. Schumaker, T. Matsuoka, V. Chvykov, F.J. Dollar, G. Kalintchenko, V. Yanovsky, A. Maksimchuk, K. Krushelnick, V.Y. Bychenkov, I.V. Glazyrin, and A.V. Karpeev, Phys. Rev. Lett. \textbf{104}, 025004 (2010).
\bibitem{bib:jrad} J.L. Martins, S.F. Martins, L.O. Silva, Proc. SPIE 7359 73590V (2009).
\bibitem{bib:thaury_prl_2013} C. Thaury, E. Guillaume, S. Corde, R. Lehe, M. Le Bouteiller, K. Ta Phuoc, X. Davoine, J. M. Rax, A. Rousse, and V. Malka, Phys. Rev. Lett. \textbf{111}, 135002 (2013).
\bibitem{bib:silva_pre_1999} L. O. Silva, R. Bingham, J.M. Dawson, W.B. Mori, Phys. Rev. E \textbf{59}, 2273 (1999); V. N. Tsytovich, \emph{Nonlinear effects in plasma} (Plenum, New York, 1970).

\bibitem{bib:esarey_pre_2002} E. Esarey, B. A. Shadwick, P. Catravas, and W.P. Leemans, Phys. Rev. E \textbf{65}, 056505 (2002).
\bibitem{bib:glinec_epl_2008} Y. Glinec, J. Faure, A. Lifschitz, J.M. Vieira, R.A. Fonseca, L.O. Silva, and V. Malka, Europhys. Lett. \textbf{81}, 64001 (2008).
\bibitem{bib:vieira_prstab_2011} J. Vieira, C.-K. Huang, W. B. Mori, and L. O. Silva, Phys. Rev. ST-AB \textbf{14} 071303 (2011).
\bibitem{bib:vieira_prl_2011} J. Vieira, S.F. Martins, V.B. Pathak, R.A. Fonseca, W.B. Mori, L.O. Silva, Phys. Rev. Lett. \textbf{106}, 225001 (2011); J. Vieira, J.L. Martins, V.B. Pathak, R.A. Fonseca, W.B. Mori, L.O. Silva, Plasma Phys. Control. Fusion, \textbf{54} 124044 (2012). 
\bibitem{bib:nemeth_prl_2008} K. N\' emeth, B. Shen, Y. Li, H. Shang, R. Crowell, K. C. Harkay, and J. R. Cary Phys. Rev. Lett. \textbf{100}, 095002 (2008).
%\bibitem{bib:yu_prl_2014} L.-L. Yu, E. Esarey, C.?B. Schroeder, J.-L. Vay, C. Benedetti, C.?G.?R. Geddes, M. Chen, W.?P. Leemans, Phys. Rev. Lett.~\textbf{112}, 125001 (2014).
\end{thebibliography}
\end{document}